# Controlling electronic properties of hexagonal manganites through aliovalent doping and thermoatmospheric history

Didrik R. Småbråten[1,2], Frida H. Danmo[1], Nikolai H. Gaukås[1], Sathya P. Singh[1], Nikola Kanas[1,3], Dennis Meier[1], Kjell Wiik[1], Mari-Ann Einarsrud[1], and Sverre M. Selbach[1,*]

[1]Department of Materials Science and Engineering, NTNU Norwegian University of Science and Technology, NO-7491 Trondheim, Norway

[2]Department of Sustainable Energy Technology, SINTEF Industry, PO Box 124 Blindern, NO-0314 Oslo, Norway.

[3]University of Novi Sad, BioSense Institute, 21000 Novi Sad, Serbia

[*]E-mail: selbach@ntnu.no

The family of hexagonal manganites is intensively studied for its multiferroicity, magnetoelectric coupling, improper ferroelectricity, functional domain walls, and topology-related scaling behaviors. It is established that these physical properties are co-determined by the cation sublattices and that aliovalent doping can readily be leveraged to modify them. The doping, however, also impacts the anion defect chemistry and semiconducting properties, which makes the system highly sensitive to the synthesis and processing conditions. Here, we study the electronic properties of YMnO$_3$ as function of aliovalent cation doping and thermoatmospheric history, combining density functional theory calculations with thermopower and thermogravimetric measurements. We show that the charge carrier concentration and transport properties can be controlled via both aliovalent cation dopants and anion defects, enabling reversible switching between *n*-type and *p*-type conductivity. This tunability is of importance for envisaged applications of hexagonal manganites in, e.g. next-generation capacitors and domain-wall nanoelectronics, or as catalysts or electrodes in fuel cells or electrolyzers. Furthermore, our approach is transferrable to other transition metal oxides, providing general guidelines for controlling their semiconducting properties.

## I. INTRODUCTION

Controlling electronic properties by donor and acceptor doping is imperative to semiconductor science and technology. Going beyond classical covalent semiconductors, such as Si and GaAs, transition metal oxides offer additional degrees of freedom for doping, utilizing oxygen defects [1,2]. Oxygen vacancies are the most common point defects and can act as electron donors, giving rise to *n*-type conductivity in systems, such as e.g. $BaTiO_3$, and $BiFeO_3$ [3]. Conversely, *p*-type conductivity can be induced by cation deficiency, e.g., Bi vacancies in $BiFeO_3$ [4], or by excess oxygen in the form of interstitials, such as in hexagonal $YMnO_3$ [5]. Importantly, compared to covalent semiconductors, transition metal oxides can tolerate many orders of magnitude higher concentration of dopants and/or point defects. Furthermore, while cation dopant concentrations are fixed during synthesis, the oxygen stoichiometry can be reversibly modified through annealing in controlled atmospheres [6]. This is possible because oxygen anions in general are mobile at much lower temperatures than the cations in metal oxides. The resulting electronic properties of doped transition metal oxides thus rely on both equilibrium thermodynamics as well as reaction and diffusion kinetics during annealing and processing.

Hexagonal manganites (h-$R$MnO$_3$, $R$ = Y, In, Sc, Ho-Lu, space group $P6_3cm$ [7,8]), are known for their rich physics, including improper ferroelectricity [9–11], frustrated magnetic order [12], topologically protected vortices [13–15], and magnetoelectric coupling [12,16]. Recently, their domain walls [13,17–19] receive broad attention as functional entities for nanoelectronics [20]. Furthermore, the hexagonal manganites exhibit an outstanding oxygen mobility [21,22] and storage properties [23], which is of interest for the field of solid state ionics [24,25]. Tailoring the local transport behavior – and the physical properties in this strongly correlated electron system in general – requires excellent control of charge carrier concentrations and distributions, which is closely linked to the cation sublattices and oxygen stoichiometry.

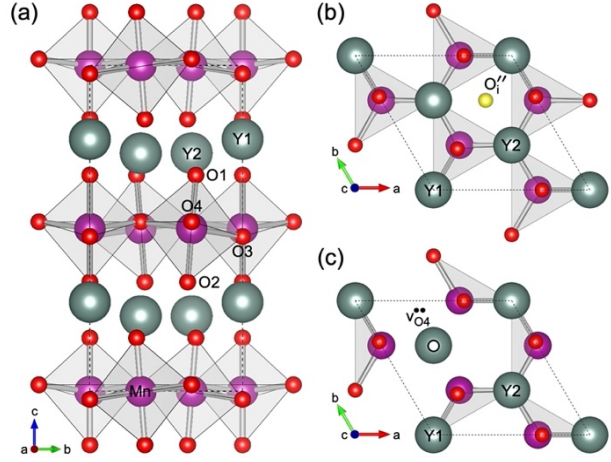

FIG. 1: (a) Crystal structure of hexagonal YMnO$_3$ in the $P6_3cm$ space group, here presented by alternating layers of Y$^{3+}$ cations and corner-sharing MnO$_5$ trigonal bipyramids. The oxygen interstitials ($O_i''$) are located in-plane and in-between three Mn$^{3+}$, illustrated in (b). The position of the energetically favored planar oxygen vacancy site O4 aligned with Y2 in the ab-plane ($V_{O4}^{\bullet\bullet}$) is shown in (c).

The $h$-$R$MnO$_3$ systems with layers of seven coordinated $R^{3+}$ cations between layers of MnO$_5$ trigonal bipyramids [FIG. 1(a)] show an outstanding chemical flexibility with respect to cation doping. In addition, they can accommodate large amounts of oxygen excess [5,24–26] through interstitials [5] and deficiency [25,27] [FIG. 1(b)], in the form of vacancies [28] [FIG. 1(c)], as well as larger defect complexes such as anti-Frenkel defect pairs [29]. This remarkable chemical flexibility and the sensitivity of the physical responses to changes in the local chemistry makes the hexagonal manganites an ideal model system for studying the correlation between aliovalent doping, thermoatmospheric history and the emergent electronic properties.

While there is already a good understanding of the coupling between point defects and domain walls [30–36], the impact of the thermoatmospheric history and how it modifies the bulk electronic properties of doped $h$-$R$MnO$_3$ is less well understood.

Here, we combine density functional theory (DFT) calculations and characterization of bulk powders to study how thermoatmospheric history influences the electronic properties in donor and acceptor-doped YMnO$_3$. We show that the conductivity is governed by the charge balance resulting from the ratios between different anion and cation defects and demonstrate switching between $n$-type and $p$-type conductivity in donor-doped samples by annealing in different atmospheres. Our results show how the electronic properties of hexagonal manganites can be reversibly controlled by annealing in different atmospheres, establishing comprehensive guidelines that are transferable to other semiconducting transition metal oxides.

## II. METHODS

### A. Computational details

The density functional theory (DFT) calculations are performed using the projector augmented wave (PAW) method implemented in VASP [37–39]. We choose 120 atoms 2x2x1 supercells as model systems, where we add one cation dopant and/or anion defect per supercell. This setting yields the stoichiometries $Y_{1-x}A_x$MnO$_{3\pm\delta}$ ($A$ = Zr$^{4+}$ or Ca$^{2+}$) for Y-substitution, and YMn$_{1-x}B_x$MnO$_{3\pm\delta}$ ($B$ = Ti$^{4+}$ or Zn$^{2+}$) for Mn-substitution, with x = $\delta$ = 1/24 (~4.2 mol%). Y (4$s$, 4$p$, 5$s$, 4$d$), Mn (3$p$, 4$s$, 3$d$), Zr (4$s$, 4$p$, 5$s$, 4$d$), Ca (3$s$, 3$p$, 4$s$), Ti (3$p$, 4$s$, 3$d$), Zn (4$s$, 3$d$) and O (2$s$, 2$p$) are treated as valence electrons. To reproduce the experimental band gap and lattice parameters of YMnO$_3$, we use the PBEsol+$U$ method [40,41] with a $U$ of 5 eV applied on Mn 3$d$ states, and a collinear frustrated antiferromagnetic ordering [42] on the Mn sublattice. The plane wave energy cutoff is set to 550 eV, and the Brillouin zone integration is performed using a Γ-centered 2x2x2 grid for the geometry optimization, which is increased to 4x4x4 for the density of states (DOS) calculations. The lattice positions for all defect cells are relaxed until the residual forces on all the atoms are below 0.005 eV Å$^{-1}$, with the lattice parameters fixed to the relaxed bulk values. The electronic band structure for pristine YMnO$_3$ is analyzed using the open-source Python package Sumo [43].

The defect formation energies of charge neutral cells are calculated as

$$E^f = E_{defect} - E_{ref} - \Sigma_i n_i \mu_i ,  \quad (1)$$

where $E_{defect}$ and $E_{ref}$ are the calculated total energies for the defect cell and the stoichiometric reference cell, and $n_i$ and $\mu_i$ are the number and chemical potential of species $i$ added to the reference cell, respectively. The defect formation energies are evaluated as a function of the chemical potential of oxygen within the DFT calculated chemical stability window of YMnO$_3$ [5] from –7.1 eV to –1.9 eV. Only neutral defect cells are evaluated as YMnO$_3$ is well-known to tolerate very high concentrations of aliovalent cation dopants [33,44–52], oxygen vacancies [25,27,28] and oxygen interstitials [5,24,25]. Thus, the dilute limit concentration regime is experimentally less relevant than for conventional semiconductors. For clarity, only the results for Y substitution on the Y2 site are included in the main text. Similar trends are obtained also for Y1-substitution (see Supp. note 1).

**B. Experimental details**

Bulk powders of donor- and acceptor-doped YMnO$_3$ are prepared using conventional solid-state reaction synthesis. Stoichiometric amounts of Y$_2$O$_3$ (99.99 %, Sigma Aldrich) and Mn$_2$O$_3$ (99 %, Sigma Aldrich), and either ZrO$_2$ (Sigma Aldrich), CaCO$_3$ (≥99 %, Merck), TiO$_2$ (99.8 %, Sigma Aldrich), or ZnO are thoroughly mixed in an agate mortar to yield a doping concentration of ~4.2 %, which corresponds to the same concentration as in our DFT calculations described above. The ZnO powder is prepared by calcining Zn(NO$_3$)$_2$ • 4H$_2$O (98.5 %, Merck) at 400°C for 15 hours. The raw powders are pressed into greenbody pellets with a diameter of 15 mm in a uniaxial press with an applied pressure of 40 MPa. All the greenbodies are sintered in air at 1300°C for 12 hours. The Ca-doped sample is subsequently sintered in nitrogen at 1400°C for 12 hours to ensure phase purity. All samples are characterized by X-ray diffraction (XRD) with a Bruker AXS D8 Focus $\theta - 2\theta$ diffractometer with Cu K$\alpha$ radiation, from 14° to 90° $2\theta$. Rietveld refinement of the diffractograms is performed with the Bruker AXS Topas academic software version 5, with initial structure parameters from ref. [9] [FIG. S3].

The thermopower measurements are performed on porous square prisms with a length of ~20 mm, a width of ~5 mm, and a height of ~3 mm, prepared by uniaxial pressing of bulk powders followed by sintering in air at 1300°C for 6 hours. The thermopower measurements are performed with a ProboStat setup (NorECs AS) in a vertical tubular furnace [53] in $N_2$ (g) and $O_2$ (g) flow for isothermals in the temperature range of 200 to 800°C upon cooling until thermal and chemical equilibria are reached.

The thermogravimetric (TG) measurements are performed with a Netzsch STA 449J Jupiter setup during heating and cooling between 50 and 900°C in oxygen, with heating and cooling rates of 1°C min$^{-1}$ and a gas flow of 30 mL min$^{-1}$. We assume oxygen stoichiometric samples ($\delta = 0$) from the mass observed at 900°C. The powders were pre-annealed *in situ* in $O_2$ (g) at 900°C and cooled as fast to 50°C at the TGA instrument allows prior to the TG measurements.

## III. RESULTS

### A. Defect chemistry model

We start by constructing a defect chemistry model for YMnO$_3$ before we calculate the electronic properties that arise due to different dopants and oxygen off-stoichiometry. The intrinsic electronic properties of YMnO$_3$ are mainly governed by two defect equilibria, that is, the electronic equilibrium,

$$K_i = np,  \qquad (2)$$

which is determined by the concentrations of the electrons ($n$) and holes ($p$), respectively, and the anti-Frenkel equilibrium

$$K_{AF} = [O_i''][V_O^{\bullet\bullet}], \qquad (3)$$

where $[O_i'']$ and $V_O^{\bullet\bullet}$ are the concentrations of oxygen interstitials and vacancies, respectively. The magnitudes of $K_i$ and $K_{AF}$ are exponentially proportional to the negative values of the electronic band gap and the anti-Frenkel formation energy, respectively. When the material is donor and/or acceptor-doped, the electroneutrality condition becomes

$$2[V_O^{\bullet\bullet}] + [D_{Y/Mn}^{\bullet}] + p = 2[O_i''] + [A_{Y/Mn}'] + n. \qquad (4)$$

Here, $[D_{Y/Mn}^{\bullet}]$ and $[A_{Y/Mn}']$ are the concentrations of the tetravalent donor and divalent acceptor dopants, respectively, located on the Y-sublattice and/or the Mn-sublattice. The electronic charge carrier concentrations $n$ and $p$ include contributions from the charge compensation of both anionic and cationic point defects.

The electroneutrality condition in Eq. (4) can naturally be shifted by changing the defect concentrations in the material system. Importantly, the resulting electronic properties can be tuned by changing the balance between the anion and cation stoichiometries, which can be achieved either by the thermoatmospheric history or the dopant content. Table I summarizes the expected majority charge carriers with respect to doping and oxygen content as described further below. Here, we assume that intermediate temperatures and high oxygen partial pressures favor oxygen interstitial formation [5,24], whereas high temperatures and low oxygen partial pressures favors oxygen vacancy formation [54]. These assumptions follow naturally from le Chatelier's principle and that the entropy part S, of Gibbs free energy, G = H – TS (where H is the enthalpy) always favors oxygen to be in the gas phase rather than being included in a regular lattice or as interstitial oxygen.

TABLE I: A qualitative summary of the expected influence cation doping and anion non-stoichiometry have on the type of conductivity in YMnO$_3$ with respect to thermal history.

| Point defects | Defect equilibrium | Conductivity | Thermal history |
|---|---|---|---|
| $O_i''$ and holes | $2[O_i''] \approx p$ | $p$-type | Low $T$, high $p(O_2)$ |
| $V_O^{\bullet\bullet}$ and electrons | $2[V_O^{\bullet\bullet}] \approx n$ | $n$-type | High $T$, low $p(O_2)$ |
| Donors | $[D_{Y/Mn}^{\bullet}] + p \approx 2[O_i''] + n$ | $n$-type or $p$-type | Medium $T$ and $p(O_2)$ |
| Acceptors | $[A_{Y/Mn}'] + n \approx 2[V_O^{\bullet\bullet}] + p$ | $p$-type or $n$-type | Medium $T$ and $p(O_2)$ |

### *Undoped YMnO$_3$*

For undoped YMnO$_3$, the electroneutrality condition in Eq. (4) is reduced to $2[V_O^{\bullet\bullet}] + p = 2[O_i''] + n$, where the resulting electronic properties depend only the ratio between the oxygen interstitial and vacancy concentrations. In oxidizing conditions, we assume that the oxygen vacancy concentration is negligible, which reduces the electroneutrality to $2[O_i''] \approx p$. Conversely, in reducing conditions, oxygen vacancies will dominate and the electroneutrality becomes $2[V_O^{\bullet\bullet}] \approx n$.

To maintain neutrality, oxygen interstitials are charge compensated by oxidizing two Mn$^{3+}$ to Mn$^{4+}$, whereas oxygen vacancies are charge compensated by reducing two Mn$^{3+}$ to Mn$^{2+}$. This is in the localized limit for electrons and holes, which is generally valid given the high effective masses of electrons and holes in hexagonal manganites (see Supp. Note 4 and FIG. S4). This can be described by the defect equations

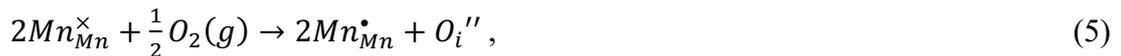

$$2Mn_{Mn}^{\times} + \tfrac{1}{2}O_2(g) \rightarrow 2Mn_{Mn}^{\bullet} + O_i'', \tag{5}$$

and

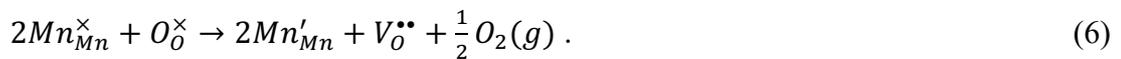

$$2Mn_{Mn}^{\times} + O_O^{\times} \rightarrow 2Mn_{Mn}' + V_O^{\bullet\bullet} + \tfrac{1}{2}O_2(g). \tag{6}$$

In sufficiently oxidizing conditions, where oxygen interstitials dominate, the system is thus expected to be *p*-type conducting. In contrast, in reducing conditions, or when oxygen vacancies dominate, we get *n*-type conductivity.

### *Donor doping*

Now that we have established the intrinsic electronic properties with oxygen excess or deficiency, we next include donor dopants in the system. For donor-doping, we expect a high affinity for forming oxygen interstitials due to electrostatics and, in turn, a negligible propensity for forming oxygen vacancies as the anti-Frenkel equilibrium will be strongly displaced towards oxygen interstitials. Hence, the electroneutrality condition in Eq. (4) can be reduced to $[D^{\bullet}_{Y/Mn}] + p = 2[O_i''] + n$, where the resulting electronic properties depend on the balance between the dopant concentration and the oxygen content. We note that for clarity, we only derive the defect chemistry model for substituting $Y^{3+}$ with $Zr^{4+}$ in the main text. The full derivation for substitution of $Mn^{3+}$ with $Ti^{4+}$ is given in Supp. Note 2.

The annealing of Zr-doped $YMnO_3$ at intermediate temperatures in varying oxygen partial pressure can be described by the general chemical equilibrium

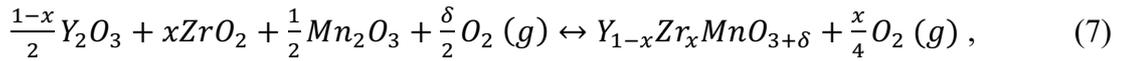

$$\frac{1-x}{2} Y_2O_3 + xZrO_2 + \frac{1}{2} Mn_2O_3 + \frac{\delta}{2} O_2 (g) \leftrightarrow Y_{1-x}Zr_xMnO_{3+\delta} + \frac{x}{4} O_2 (g), \qquad (7)$$

where *x* is the dopant concentration and $\delta$ is the oxygen hyper-stoichiometry. In sufficiently low $p(O_2)$, we can assume an oxygen stoichiometry equal to $\delta = 0$. Microscopically, Eq. (7) becomes

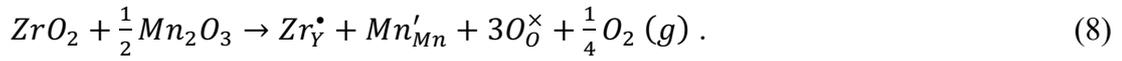

$$ZrO_2 + \frac{1}{2} Mn_2O_3 \rightarrow Zr^{\bullet}_Y + Mn'_{Mn} + 3O^{\times}_O + \frac{1}{4} O_2 (g). \qquad (8)$$

Here, the donor dopant $Zr^{\bullet}_Y$ is charge compensated by reducing one $Mn^{3+}$ to $Mn^{2+}$, which renders the material *n*-type conducting, as expected for donor-doping. With increasing oxygen partial pressure, oxygen interstitials are progressively favored. Assuming an oxygen interstitial concentration according to $[O_i''] = \frac{1}{2}[Zr_Y^\bullet]$, the oxygen interstitials are ionically charge compensated by the donor dopant. This is termed the neutral condition, described by

$$ZrO_2 + \frac{1}{2}Mn_2O_3 \rightarrow Zr_Y^\bullet + Mn_{Mn}^\times + 3O_O^\times + \frac{1}{2}O_i''. \tag{9}$$

For this specific condition, the material is expected to be neither *n*-type nor *p*-type, which should give an electronic conductivity similar to that of stoichiometric YMnO$_3$ as the electronic charge carrier concentration remains unchanged. Further increasing the oxygen content exceeding the neutral conditions, the excess oxygen interstitials are charge compensated by oxidizing Mn$^{3+}$ to Mn$^{4+}$ according to Eq. (5), which renders the material *p*-type. In our explicit DFT calculations below, we add one oxygen interstitial per donor dopant, corresponding to the condition $[O_i''] = [Zr_Y^\bullet]$. For this situation, eq. (7) simplifies to:

$$ZrO_2 + \frac{1}{2}Mn_2O_3 + \frac{1}{4}O_2(g) \rightarrow Zr_Y^\bullet + Mn_{Mn}^\bullet + 3O_O^\times + O_i''. \tag{10}$$

The net result is one Mn$^{3+}$ being oxidized to Mn$^{4+}$, which renders the material *p*-type conducting.

Similarly, for Ti-doping, when the oxygen content is $\delta = 0$, we get the following formation reaction equation:

$$\frac{1}{2}Y_2O_3 + TiO_2 + Mn_{Mn}^\times \rightarrow Y_Y^\times + Ti_{Mn}^\bullet + Mn_{Mn}' + 3O_O^\times + \frac{1}{4}O_2(g), \tag{11}$$

which for neutral conditions, i.e., $[O_i''] = \frac{1}{2}[Ti_{Mn}^\bullet]$ simplifies to

$$\frac{1}{2}Y_2O_3 + TiO_2 \rightarrow Y_Y^\times + Ti_{Mn}^\bullet + 3O_O^\times + \frac{1}{2}O_i''. \tag{12}$$

In our DFT calculations we choose the condition $[O_i''] = [Ti_{Mn}^\bullet]$, which gives the following formation reaction:

$$\frac{1}{2}Y_2O_3 + TiO_2 + Mn_{Mn}^\times + \frac{1}{4}O_2(g) \rightarrow Y_Y^\times + Ti_{Mn}^\bullet + Mn_{Mn}^\bullet + 3O_O^\times + O_i''. \tag{13}$$

To summarize, when donor-doped YMnO$_3$ is annealed in increasing $p(O_2)$, we go from *n*-type conductivity in low $p(O_2)$ (Eqs. (8) and (11)), through a minimum in the conductivity for intermediate $p(O_2)$ (Eqs. (9) and (12)), to $p$-type conductivity in high $p(O_2)$ (Eqs. (10) and (13)).

## *Acceptor doping*

Next, we address acceptor-doping with respect to varying oxygen content. In contrast to the donor-doping described above, we expect a high affinity for forming oxygen vacancies and a negligible affinity for forming oxygen interstitials as the anti-Frenkel equilibrium is shifted towards oxygen vacancies. Hence, the electroneutrality condition in Eq. (4) can be reduced to $2[V_O^{\bullet\bullet}] + p = [A'_{Y/Mn}] + n$, showing that the electronic properties depend on the ratio between the dopant concentration and oxygen content. For clarity, we only derive the defect chemistry model for substituting $Y^{3+}$ with $Ca^{2+}$ in the main text. The full derivation for substitution of $Mn^{3+}$ with $Zn^{2+}$ is explained in Supp. Note 3.

The annealing of Ca-doped YMnO$_3$ at high temperatures in varying oxygen partial pressure can be described by the general chemical equilibrium

$$\frac{1-x}{2}Y_2O_3 + xCaO + \frac{1}{2}Mn_2O_3 + \frac{x}{2}O_2(g) \leftrightarrow Y_{1-x}Ca_xMnO_{3-\delta} + \frac{\delta}{4}O_2(g), \quad (14)$$

where $x$ is the dopant concentration and $\delta$ is the oxygen hypo-stoichiometry. In sufficiently high $p(O_2)$, we can assume an oxygen stoichiometry with $\delta = 0$. In this case, Eq. (14) becomes

$$CaO + \frac{1}{2}Mn_2O_3 + \frac{1}{4}O_2(g) \rightarrow Ca'_Y + Mn^{\bullet}_{Mn} + 3O_O^{\times}. \quad (15)$$

Here, the acceptor dopant $Ca'_Y$ is charge compensated by oxidizing one $Mn^{3+}$ to $Mn^{4+}$ which renders the material $p$-type conducting as expected for acceptor-doping. By lowering the oxygen partial pressure, oxygen vacancies start to form. Assuming neutral conditions, corresponding to $[V_O^{\bullet\bullet}] = \frac{1}{2}[Ca'_Y]$, the oxygen vacancies are ionically charge compensated by the acceptor dopant according to

$$CaO + \frac{1}{2}Mn_2O_3 \rightarrow Ca'_Y + Mn^{\times}_{Mn} + \frac{5}{2}O_O^{\times} + \frac{1}{2}V_O^{\bullet\bullet}. \quad (16)$$

For these specific conditions, the material is expected to be neither *n*-type nor *p*-type dominating, which should give an electronic conductivity similar to that of stoichiometric YMnO$_3$, analogous to the scenario described for donor-doping above. Further decreasing the oxygen content below neutral conditions, the excess oxygen vacancies are charge compensated by reducing Mn$^{3+}$ to Mn$^{2+}$ according to Eq. (6), rendering the material *n*-type conducting. In our explicit DFT calculations below, we add one oxygen vacancy per acceptor dopant, corresponding to the conditions $[V_O^{\bullet\bullet}] = [Ca_Y']$. Eq. (14) can in this case be simplified to:

$$CaO + \frac{1}{2}Mn_2O_3 \rightarrow Ca_Y' + Mn_{Mn}' + 2O_O^\times + V_O^{\bullet\bullet} + \frac{1}{4}O_2\,(g)\,, \qquad (17)$$

where the net result is one Mn$^{3+}$ being reduced to Mn$^{2+}$, which renders the material *n*-type conducting.

Similarly, for Zn-doping, we have for an oxygen content of $\delta = 0$:

$$\frac{1}{2}Y_2O_3 + ZnO + Mn_{Mn}^\times + \frac{1}{4}O_2\,(g) \rightarrow Y_Y^\times + Zn_{Mn}' + Mn_{Mn}^\bullet + 3O_O^\times\,, \qquad (18)$$

for neutral conditions, i.e., $[V_O^{\bullet\bullet}] = \frac{1}{2}[Zn_{Mn}']$

$$\frac{1}{2}Y_2O_3 + ZnO \rightarrow Y_Y^\times + Zn_{Mn}' + \frac{5}{2}O_O^\times + \frac{1}{2}V_O^{\bullet\bullet}\,, \qquad (19)$$

and for the DFT calculations, $[V_O^{\bullet\bullet}] = [Zn_{Mn}']$, which gives the following reaction:

$$\frac{1}{2}Y_2O_3 + ZnO + Mn_{Mn}^\times \rightarrow Y_Y^\times + Zn_{Mn}' + Mn_{Mn}' + 2O_O^\times + V_O^{\bullet\bullet} + \frac{1}{4}O_2\,(g)\,. \qquad (20).$$

To summarize, we expect that when we anneal acceptor-doped YMnO$_3$ with decreasing *p*(O$_2$), we go from *p*-type conductivity in high *p*(O$_2$) (Eqs. (15) and (18)), through a minimum in the conductivity for intermediate *p*(O$_2$) (Eqs. (16) and (19)), to *n*-type conductivity in low *p*(O$_2$) (Eqs. (17) and (20)). A graphical summary of the interplay between aliovalent cation dopants, oxygen defects and *n*- or *p*-type conductivity is given in FIG. 2.

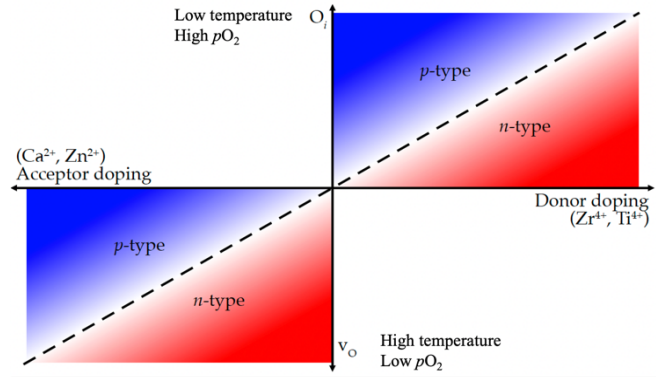

FIG. 2. Schematic illustration of how *n*- and *p*-type conductivity depends on donor and acceptor doping and processing conditions in the form thermoatmospheric history. The horizontal line represents donor dopant concentration to the right and acceptor dopant concentration to the left. The vertical line represents oxygen interstitial concentration upwards, and oxygen vacancy concentration downwards. Blue fill corresponds to increasing *p*-type conductivity and red fill corresponds to increasing *n*-type conductivity. The diagonal dashed line represents the charge neutral condition, assuming $[O_i''] = \frac{1}{2}[D^\bullet_{Y/Mn}]$ and $[V_O^{\bullet\bullet}] = \frac{1}{2}[A'_{Y/Mn}]$ for donor doping and acceptor doping, respectively.

## B. Computational predictions

### *Electronic properties*

Following our defect chemistry model, we use DFT to calculate the electronic properties for the different selected chemical compositions of interest. First, we focus on the electronic properties resulting from oxygen interstitials and oxygen vacancies in undoped YMnO$_3$. The calculated electronic density of states (DOS) for stoichiometric YMnO$_3$ is shown in FIG. 3(a). The valence band consists mainly of O ($2p_x$, $2p_y$) states, whereas the conduction band is dominated by empty Mn $3d_{z^2}$ states, in agreement with ref. [7]. The average Bader charge for the Mn cations is equal to +1.71 and the average Mn magnetic moment is 3.7 $\mu$B. In the following, we will use relative changes in the calculated Bader charges and magnetic moments of Mn as an indication of oxidation or reduction of $Mn^{3+}$, where increased Bader charges and decreased magnetic moments indicate oxidation from $Mn^{3+}$ to $Mn^{4+}$, whereas reduction from $Mn^{3+}$ to $Mn^{2+}$ is indicated by decreased Bader charges and increased magnetic moments.

The resulting DOS for $O_i''$ in undoped $YMnO_3$ is shown in FIG. 3(b). We observe a non-bonding state in the band gap consisting of mainly of empty Mn ($3d_{x^2-y^2}$, $3d_{xy}$) and $O_i''$ ($2p_x$, $2p_y$). Integrating the DOS between the valence band maximum and the conduction band minimum corresponds to two holes. The corresponding Mn-$O_i''$ binding states are found at the bottom of the valence band. The two Mn closest to the $O_i''$ [FIG. 1(b)] show increased Bader charges of +1.89, accompanied by lower magnetic moments of 3.1 $\mu$B. This corresponds to the formal oxidation of two $Mn^{3+}$ to $Mn^{4+}$ according to Eq. (5), rendering the material $p$-type conducting consistent with ref. [5].

The calculated DOS for the energetically most favored planar O4 oxygen vacancy $V_{O4}^{\bullet\bullet}$ [28] is shown in FIG. 3(h). Formation of $V_{O4}^{\bullet\bullet}$ leads to occupied Mn $3d_{z^2}$ states in the band gap, which integrate to two electrons from the DOS. Bader charge analysis reveals two Mn in the vicinity of the oxygen vacancy with decreased Bader charges of +1.37 and increased magnetic moments of 4.4 $\mu$B. This signifies a formal reduction of two $Mn^{3+}$ to $Mn^{2+}$ according to Eq. (6) and $n$-type in agreement with ref. [28].

The changes that arise in the DOS of donor-doped $YMnO_3$ with and without oxygen interstitials are displayed in FIG. 3(c-e). Compared to stoichiometric $YMnO_3$ in FIG. 3(a), donor-doping with $Zr^{4+}$ [FIG. 3(c)] and $Ti^{4+}$ [FIG. 3(e)] raises the Fermi level into the intrinsic conduction band. Integrating the occupied states in the conduction band for each of the donor dopants gives one electron, which occupies the initially empty Mn $3d_{z^2}$ states. Zr-doping results in a net reduction in the Bader charge of the Mn sublattice of 0.24 relative to pristine $YMnO_3$, which is mostly localized on one Mn in the vicinity of the dopant with a decreased Bader charge of +1.59, accompanied by an increased magnetic moment of 4.1 $\mu$B. Similarly,

Ti-doping results in a net decreased Bader charge in the Mn sublattice of –0.33. However, here the charge compensated electrons are delocalized over Mn in the same plane as the dopant. These 'in-plane' Mn also show a small increase in their magnetic moments equal to 3.8 $\mu$B. The results indicate a reduction of $Mn^{3+}$ to $Mn^{2+}$ by donor-doping with $Zr^{4+}$ or $Ti^{4+}$, which renders the material $n$-type according to Eqs. (8) and (11).

The resulting DOS from adding one $O_i''$ to Zr-doped material is shown in FIG. 3(d). As a consequence, the Fermi level is pushed just below the intrinsic valence band edge. One Mn shows an increased Bader charge of 1.91 and a corresponding decreased magnetic moment of 3.1 $\mu$B. Hence, the net result of having one Zr-dopant and one oxygen interstitial is an oxidation of one $Mn^{3+}$ to $Mn^{4+}$. Similarly, introducing $O_i''$ in the Ti-doped system pushes the Fermi level back to the intrinsic valence band edge as shown in FIG. 3(f). One Mn shows an increased Bader charge of +1.90 and a decreased magnetic moment of 3.0 $\mu$B, corresponding to an oxidation of one $Mn^{3+}$ to $Mn^{4+}$. As for $O_i''$ in undoped $YMnO_3$, we observe the characteristic binding states between $Mn^{4+}$ and $O_i''$ at the bottom of the valence band also for donor-doped $YMnO_3$. These results suggest that at sufficiently high $O_i''$ concentration, donor-doped $YMnO_3$ becomes $p$-type conducting due to a net oxidation of $Mn^{3+}$ to $Mn^{4+}$ according to Eqs. (10) and (13).

We continue with the analysis of the electronic properties of acceptor-doped $YMnO_3$ with and without oxygen vacancies, as illustrated by the changes in the DOS in the right panels in FIG. 3. For both Ca-doping and Zn-doping in FIG. 3(i) and (k), respectively, an unoccupied state consisting mainly of Mn $3d_{x^2-y^2}$ and $3d_{xy}$ emerges ~0.1 eV above the valence band edge, equal to one hole from the integrated DOS. Both acceptor dopants give a subtle net increase in the calculated Bader charges in the Mn sublattice of 0.05 and 0.02 for Ca-doping and Zn-doping, respectively, where the charge compensating holes are delocalized in the Mn layer closest to the dopants. For the Ca-doped system, two Mn show a small increase in their calculated Bader charges to +1.75 and magnetic moments decreased to 3.6 $\mu$B, indicating oxidation of $Mn^{3+}$ to $Mn^{4+}$. Zn-doping gives one completely delocalized hole associated with Mn in the same plane as the dopant. These results indicate that Ca-doping and Zn-doping renders the material *p*-type conducting, according to Eqs. (14) and (17), respectively.

FIG. 3(j) and (l) show the resulting DOS when adding $V_O^{\bullet\bullet}$ to the Ca-doped and Zn-doped system, respectively. In both cases, an occupied defect state in the band gap of mainly Mn $3d_{z^2}$ character is observed, equal to one electron from the integrated DOS. The charge compensating electrons are mainly localized on one Mn with a Bader charge decreased to +1.40 and +1.41 in the Ca- and Zn-doped systems, respectively. Similarly, the same Mn show increased magnetic moments of 4.3 $\mu$B and 4.4 $\mu$B, respectively. These results show that for sufficiently high $V_O^{\bullet\bullet}$ concentration, acceptor-doped $YMnO_3$ becomes *n*-type conducting by a net reduction of $Mn^{3+}$ to $Mn^{2+}$ according to Eqs. (16) and (20).

In summary, we find that the electronic properties in doped $YMnO_3$ are determined by two equivalently important factors, that is, (i) the doping concentration and (ii) the anion defect chemistry. The combined effect of (i) and (ii) is crucial to understand the diverse electronic responses reported for hexagonal manganites and emphasizes the importance of the growth parameters. Most importantly, with respect to the engineering of the specific *p*- and *n*-type characteristics, the results indicate that it is possible to control, or fine-tune, the semiconducting behavior of both undoped and doped samples via the thermoatmospheric history.

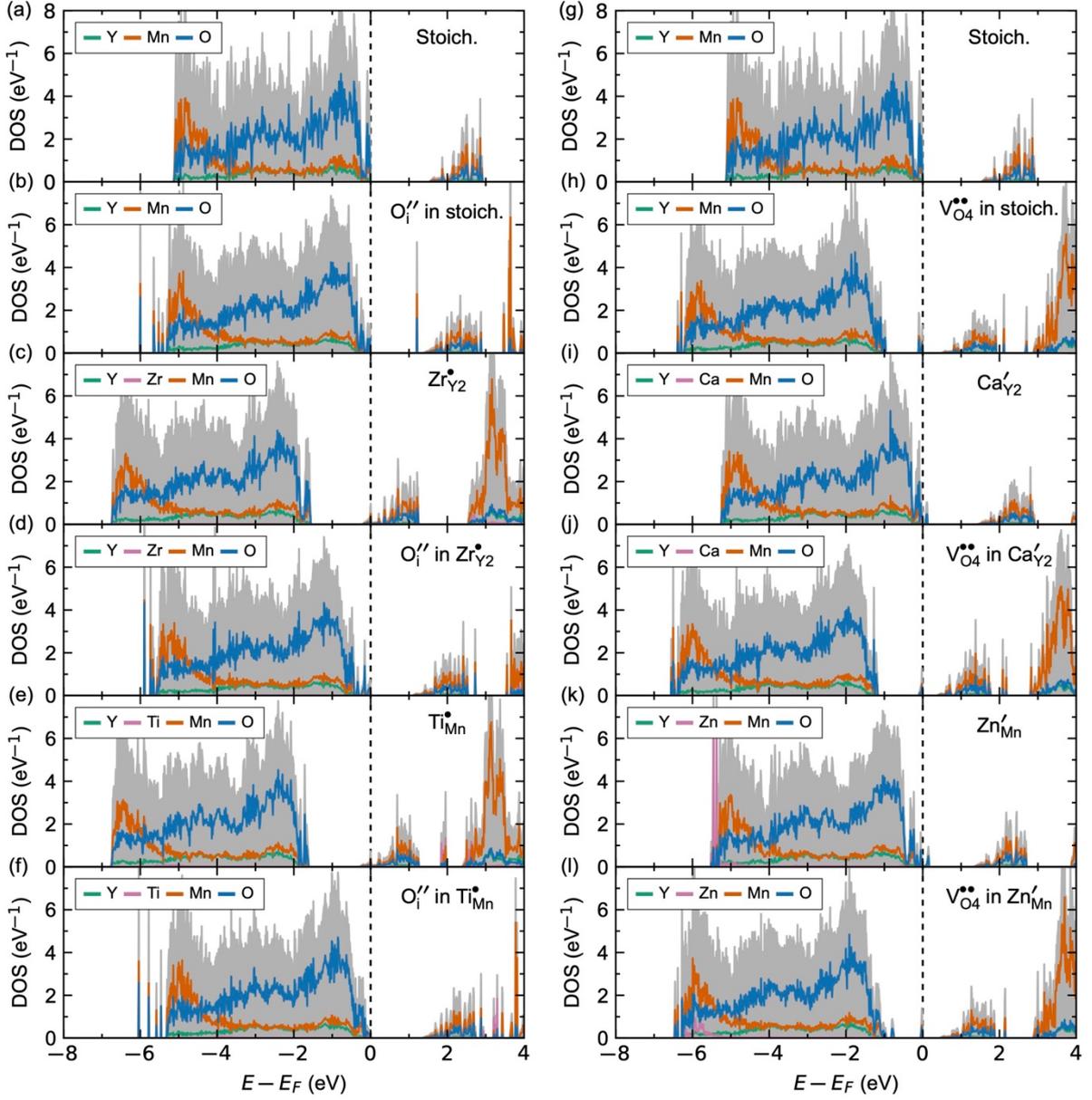

FIG. 3: Calculated DOS for 120 atoms 2x2x1 supercells of (a,g) stoichiometric $YMnO_3$, (b) $O_i''$ in undoped $YMnO_3$, (c) Zr-doped $YMnO_3$, (d) $O_i''$ in Zr-doped $YMnO_3$, (e) Ti-doped $YMnO_3$, (f) $O_i''$ in Ti-doped $YMnO_3$, (h) $V_O^{\bullet\bullet}$ in undoped $YMnO_3$, (i) Ca-doped $YMnO_3$, (j) $V_O^{\bullet\bullet}$ in Ca-doped $YMnO_3$, (k) Zn-doped $YMnO_3$, and (l) $V_O^{\bullet\bullet}$ in Zn-doped $YMnO_3$. The left panels illustrate the changes in the DOS for donor-doping with and without $O_i''$, while the right panels illustrate the changes in the DOS for acceptor-doping with and without $V_O^{\bullet\bullet}$. The dashed vertical lines illustrate the position of the Fermi level relative to the intrinsic band edges for the deferent supercells.

*Defect energetics*

After discussing the impact of anion oxygen point defects on the electronic carrier type and concentration, we next investigate the corresponding energetics for the anion point defects with respect to cation doping. Simple electrostatic considerations suggest that negatively charged oxygen interstitials form more easily in donor-doped systems, where they would charge balance and be attracted to dopants with a positive charge relative to the undoped material. Correspondingly, positively charged oxygen vacancies are energetically favored in acceptor-doped systems, where the dopants have a negative charge compared to the stoichiometric material.

The calculated defect formation energies for $O_i''$ and $V_O^{\bullet\bullet}$ in undoped and doped YMnO$_3$ as a function of the oxygen chemical potential are shown in FIG. 4(a) and (b), respectively. The formation energy for $O_i''$ is significantly reduced for the donor-doped systems as compared to the undoped system, with a formation energy lowering of 0.87 eV and 1.15 eV for Zr- and Ti-doping, respectively. The difference in formation energy lowering may be partly attributed to $O_i''$ being situated in the same plane as Ti which substitutes Mn, making the dipole energy smaller than in the case of Zr which substitutes Y. This reduction in formation energy can be rationalized from electronic structure considerations as donor doping alone leads to occupation of the higher-energy Mn $3d_{z^2}$ states. Incorporation of $O_i''$ provides vacant O $2p$ states which accept the donated electrons from the dopants at a much lower energy. To highlight the dominant electronic contribution to the lowering of the $O_i''$ formation energy, we inspect the DOS in FIG. 3(a)-(f) and find that $E_F$ is lowered by 1.36 eV and 1.64 eV upon adding an $O_i''$ to Zr-doped and Ti-doped YMnO$_3$, respectively. The calculated energy lowering for $O_i''$ formation in donor-doped YMnO$_3$ compared to undoped YMnO$_3$ from both the defect formation energies and $E_F$ are summarized in Table II.

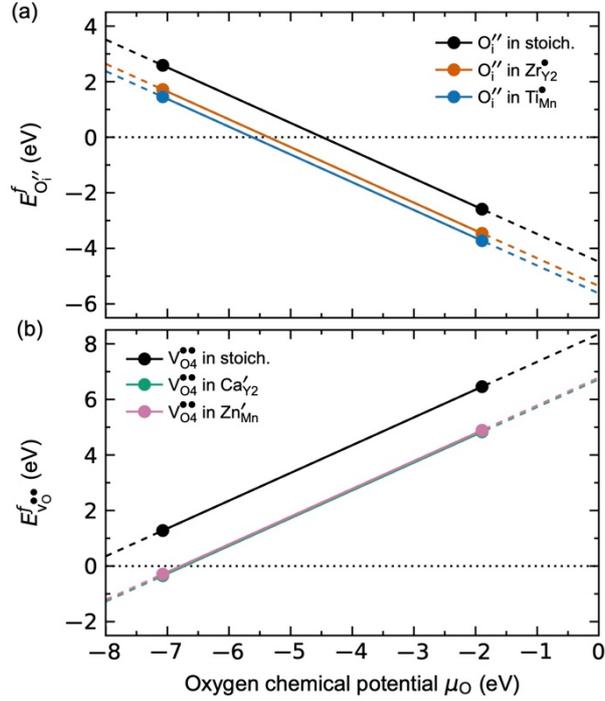

FIG. 4: A comparison of the calculated defect formation energies in 120 atoms 2x2x1 supercells as a function of the oxygen chemical potential for (a) $O_i''$ in undoped and donor-doped YMnO$_{3+\delta}$, and (b) $V_O^{\bullet\bullet}$ in undoped and acceptor-doped YMnO$_{3-\delta}$. Note that only neutral defect cells are evaluated since YMnO$_3$ is well-known to tolerate very high concentrations of the defects studied here.

FIG. 4(b) shows the calculated formation energies of $V_O^{\bullet\bullet}$ in undoped and acceptor-doped YMnO$_3$ as a function of the oxygen chemical potential. In analogy with the donor-doped systems described above, acceptor-doping gives a significant reduction in the defect formation energy of oxygen vacancies, with an energy lowering compared to undoped YMnO$_3$ of 1.63 eV and 1.57 eV for Ca- and Zn-doping, respectively. Again, this energy lowering can partially be reasoned from the changes in the DOS in FIG. 3(g)-(l). The formation of $V_O^{\bullet\bullet}$ in undoped YMnO$_3$ results in two occupied electronic states in the band gap, accompanied by an increase in the Fermi energy of 1.02 eV. For $V_O^{\bullet\bullet}$ in the acceptor-doped systems, only one electron is occupying a state within the band gap. The resulting increases in the Fermi energies with oxygen vacancy formation are 1.14 eV and 0.85 eV for the Ca- and the Zn-doped systems, respectively. Assuming that the oxygen vacancy formation energy is proportional to the sum of the increased Fermi energy per electron occupied in the band gap, the corresponding energy lowering from the DOS becomes 0.90 eV and 1.19 eV for forming $V_O^{\bullet\bullet}$ in Ca- and Zn-doped YMnO$_3$, respectively. The calculated energy for oxygen vacancy formation in acceptor-doped YMnO$_3$ as compared to undoped YMnO$_3$ from both the defect formation energies and the DOS are summarized in Table II.

Note that the differences between the calculated energy gains from the formation energies and the changes in the Fermi energy are not identical due to additional effects of elastic strain and perturbations of the Coulombic attraction forces in this predominantly ionic material.

To summarize this section, we find a strong affinity for forming oxygen interstitials in donor-doped YMnO$_3$, and for forming oxygen vacancies in acceptor-doped YMnO$_3$, compared to undoped YMnO$_3$. Since the formation energy of $O_i''$ is significantly reduced upon donor-doping, we expect that as-prepared donor-doped samples in air should show significant oxygen hyper-stoichiometry compared to as-prepared undoped YMnO$_3$ because of the enhanced enthalpy stabilization of interstitials that is expected to compensate the entropy cost at a lower temperature. A consequence is that compared to undoped YMnO$_3$, oxygen interstitials can form at higher temperatures in donor-doped YMnO$_3$ during cooling. Conversely, since the formation energy of oxygen vacancies is significantly lowered by acceptor-doping, we expect that oxygen vacancies will form at lower temperatures in acceptor-doped samples compared to undoped.

TABLE II: Calculated energy gain by forming oxygen interstitials ($O_i''$) in donor-doped YMnO$_3$, and by forming oxygen vacancies ($V_O^{\bullet\bullet}$) in acceptor-doped YMnO$_3$, as compared to undoped YMnO$_3$. The first column gives the energy gain from the calculated formation energies ($E_{def}^f$) in FIG. 4, and the second column from the changes in the Fermi level ($E_F$) from the calculated DOS in FIG. 3.

| System | Energy gain (eV) | |
|---|---|---|
| | From $E_{def}^f$ | From $E_F$ |
| $O_i''$ and $Zr_{Y2}^{\bullet}$ | 0.87 | 1.36 |
| $O_i''$ and $Ti_{Mn}^{\bullet}$ | 1.14 | 1.64 |
| $V_O^{\bullet\bullet}$ and $Ca_{Y2}'$ | 1.63 | 0.90 |
| $V_O^{\bullet\bullet}$ and $Zn_{Mn}'$ | 1.57 | 1.19 |

### C. Experimental observations

To corroborate our defect chemistry model for donor- and acceptor-doped YMnO$_3$ and confirm the strong correlation between semiconducting properties and thermoatmospheric history, we perform thermopower and thermogravimetric measurements on donor- and acceptor-doped bulk polycrystals.

*Thermopower measurements*

The type of conductivity with respect to doping and thermoatmospheric history can be determined by thermopower measurements, where a positive (negative) sign of the Seebeck coefficient indicates that holes (electrons) are dominating the electronic transport properties. We have previously reported thermopower measurements of undoped YMnO$_3$ polycrystals [5], where the Seebeck coefficient was observed to change from positive to negative upon heating from 350 to 400°C in $N_2$ atmosphere.

The measured Seebeck coefficient, $S$, as a function of temperature for polycrystalline samples of donor-doped $YMnO_3$ in $N_2$ (g) and $O_2$ (g) flow are shown in FIG. 5(a)-(b). A negative Seebeck coefficient in $N_2$ (g) is observed for all temperatures, indicating *n*-type conductivity. This is expected for donor-doping and in agreement with our defect chemistry model (Eqs. 8 and 11) and DFT predictions in oxygen-poor conditions as described above. When we switch to $O_2$ (g), both samples show *n*-type behavior at temperatures exceeding 450-500°C. However, upon cooling below 400-450°C, the Seebeck coefficient changes from negative to positive and remains positive upon further cooling. This behavior reflects a switch from *n*-type behavior to *p*-type conductivity at intermediate temperatures in $O_2$ (g). Recalling that oxygen interstitials in undoped $YMnO_3$ are typically incorporated at temperatures in the range of 250-350°C [5,24], this gives a strong indication that the switch from *n*-type to *p*-type conductivity below 400-450°C is an effect of incorporation of sufficiently large amounts of oxygen interstitials, in agreement with the defect chemistry model (Eqs. 10 and 13).

For the acceptor-doped samples in FIG. 5(c)-(d), we see no clear trend in the measured Seebeck coefficient with respect to the atmosphere during thermal treatment. Both samples show positive Seebeck coefficients (*p*-type conductivity) for all temperatures in both $N_2$ (g) and $O_2$ (g). At 800°C in $N_2$ (g), both samples show a small reduction in the Seebeck coefficient, indicating that our experimental setup could not reach sufficiently high temperatures or sufficiently low $p(O_2)$ to switch to *n*-type conductivity.

To summarize the thermopower measurements, we see pronounced changes in the Seebeck coefficient for the donor-doped samples, where a change from negative to positive values at intermediate temperatures when changing from $N_2$ (g) to $O_2$ (g) is observed, showing a switch from *n*-type to *p*-type conductivity with increasing oxidizing conditions. Conversely, a small dip in the Seebeck coefficient for the acceptor-doped samples at higher temperatures is observed, which indicates that even higher temperatures or lower $p(O_2)$ is required to switch the acceptor-doped samples to *n*-type conductivity.

*Thermogravimetric analysis*

To explicitly determine the changes in oxygen content in the samples as inferred by the Seebeck coefficient measurements above, we next perform thermogravimetric (TG) measurements of the bulk powders during heating and cooling in $O_2$ (g). The resulting oxygen stoichiometry (3+$\delta$) as a function of temperature for the donor-doped and acceptor-doped powders are shown in FIG. 6(a) and (b), respectively.

During heating the samples in $O_2$ (g), we observe a constant mass below ~150°C. In the temperature region of ~150-300°C, we see a steady increase in the mass with a maxiumum in the oxygen stoichiometry at 3.08, 3.04, 3.01, 3.03 for the Zr-, Ti-, Ca-, and Zn-doped samples, respectively. For heating above ~300°C, we observe a steady decrease in the masses that slowly converge towards a constant value at elevated temperatures (~900°C). This constant value is assumed to correspond to a stoichiometric oxygen content where $\delta = 0$. The mass change during the heating cycles indicates that all four samples show residual oxygen excess, originating either from the synthesis and/or from the pre-annealing step, that is released upon heating. We note that the two acceptor-doped samples show initially lower oxygen excess than the donor-doped samples, as expected from electrostatics and from the computational predictions described above.

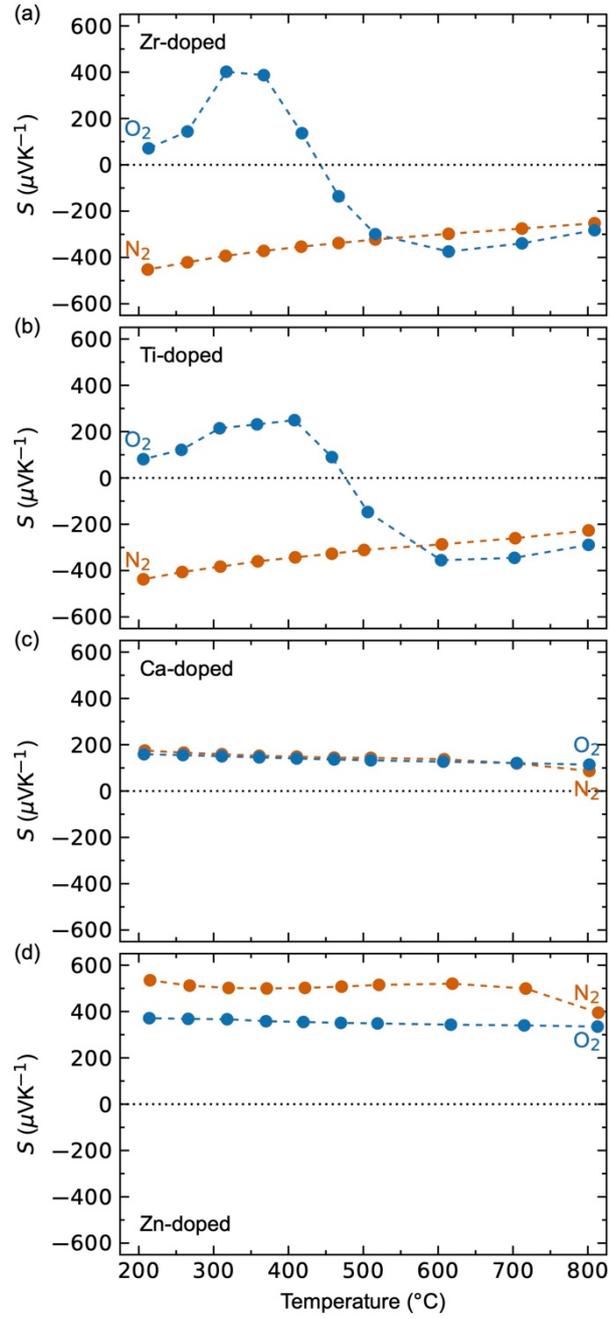

FIG. 5: Measured Seebeck coefficient ($S$) as a function of temperature in $N_2$ (g) and $O_2$ (g), for (a) Zr-doped, (b) Ti-doped, (c) Ca-doped, and (d) Zn-doped $YMnO_3$ polycrystalline samples.

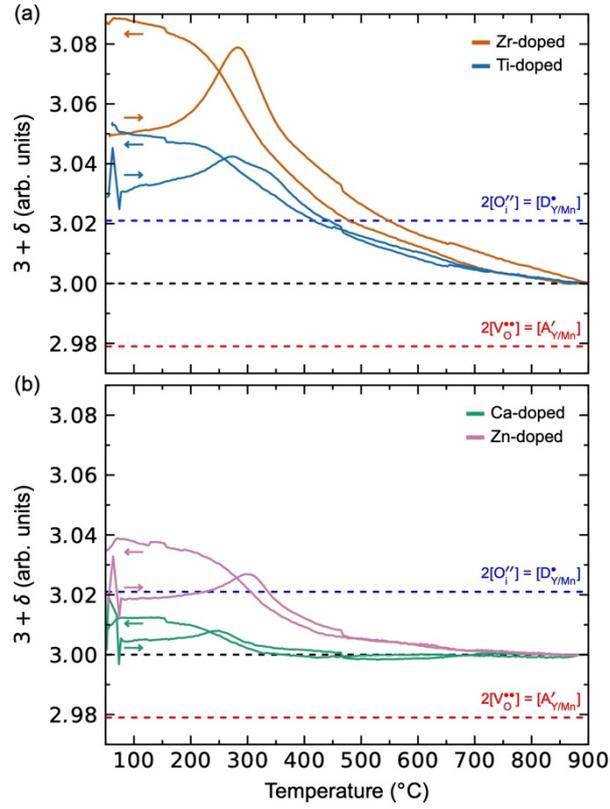

FIG. 6: Estimated oxygen content $(3 + \delta)$ in (a) donor-doped, and (b) acceptor-doped YMnO$_{3+\delta}$ bulk powders from TG measurements during heating and cooling in O$_2$ (g). The stoichiometric oxygen content ($\delta = 0$) is approximated from the measured mass at 900°C. The neutral conditions $2[O_i''] = [D_{Y/Mn}^\bullet]$ and $2[V_O^{\bullet\bullet}] = [A_{Y/Mn}']$ are marked by the dashed horizontal blue and red lines, respectively. Heating and cooling directions are illustrated in arrows.

When the donor-doped samples are subsequently cooled in O$_2$ (g) [FIG. 6(a)], we observe a continuous increase in the mass until the mass flattens out to a non-stoichiometric $(3+\delta)$ plateau equal to 3.09 and 3.05 for the Zr- and Ti-doped samples, respectively. These results indicate that oxygen interstitials are incorporated into the samples upon cooling in air, similar to what is reported in undoped YMnO$_3$ and in $(R1_{1-x}R2_x)$MnO$_3$ solid solutions [5,24]. The increases in the masses coincide with the region around 350 to 500°C for the switching from negative to positive Seebeck coefficient in FIG. 5(a)-(b), which further implies that the observed changes in the electronic properties are mainly caused by incorporation of sufficiently large amounts of oxygen excess through oxygen interstitials. The acceptor-doped samples also show a non-stoichiometric (3+δ) plateau after cooling in $O_2$ (g) equal to 3.01 and 3.04 for the Ca- and Zn-doped samples, respectively.

## IV. DISCUSSION

The formation energy of oxygen vacancies in the hexagonal manganites is relatively high[28], hence reducing the samples in our experimental setup is difficult as the $N_2$ (g) flow has a finite $p(O_2)$, estimated to $10^{-4}$-$10^{-5}$ bar. To significantly reduce the samples, one would either require strongly reducing atmospheres and sufficiently elevated temperatures, or to use other methods such as, e.g., topotactic reduction [27]. The more positive Seebeck coefficient observed for the Zn-doped sample in $N_2$ (g) compared to $O_2$ (g) is counter-intuitive as more reducing atmospheres should favor less positive *S*-values. This is not fully understood, but could be related to a composite-like effect due to inhomogeneous reduction of the sample, as observed in e.g., acceptor-doped $BiFe_{1-x}Co_xFeO_3$ [55].

Although our TGA results show oxygen hyperstoichiometry in the donor-doped samples after cooling in $O_2$ (g), the reported values are significantly lower than what is reported for undoped $YMnO_3$ [24]. As the maximum amount of oxygen excess is mainly governed by the limit at which no more oxygen can be incorporated without the unlikely oxidation of $Mn^{4+}$ to $Mn^{5+}$, [5] we expect the inherently reduced donor-doped samples to have a much higher oxygen storage capacity compared to undoped samples, as reported for Zr-doped [26] and Ti-doped [21,22,48] materials. We attribute our low oxygen content to poor kinetics due to long diffusion distance in bulk-sized powder particles [6], and we expect a significant increase in the TGA measured storage capacity for nanoparticles and for lower cooling rates [24]. This is further corroborated by the measured Seebeck coefficients in Fig 5(a)-(b), which clearly suggest that we have a high concentration of oxygen interstitials in the donor-doped samples when they are kept at intermediate temperatures in $O_2$ (g) for a sufficient amount of time. In contrast, the acceptor-doped samples, are expected to have a much lower oxygen storage capacity, as observed here, due to the inherent oxidation from the doping.

## V. CONCLUSIONS

We have determined the important role of thermoatmospheric history and doping on the electronic properties of hexagonal manganites. Our defect chemistry model shows that the charge carrier concentration and corresponding conductivity depend on the ratio between the concentration of dopants and anion defects. This relation is further confirmed by DFT calculations, where we find that donor (acceptor) doped systems with inherent $n$-type ($p$-type) conductivity can be switched to $p$-type ($n$-type) conducting with sufficient amounts of oxygen interstitials (vacancies). Finally, our theoretical predictions are corroborated by Seebeck coefficient and TG measurements on bulk samples, which show that donor-doped materials can be switched between $n$-type and $p$-type conductivity by annealing in different atmospheres. These results clarify the electronic properties of the hexagonal manganites and explains their diverse conduction behaviors discussed in the literature.

The implications of our results extend far beyond hexagonal manganites and demonstrate how aliovalently doped semiconducting transition metal oxides can be reversibly made $n$-type or $p$-type conducting through annealing in controlled temperature and atmosphere. Candidate materials could be perovskites which are prone to both cation and oxygen vacancies, such as $BiFeO_3$ [3,4], $PbTiO_3$ [56] and $LiNbO_3$ [57]. For example, rewritable $p$-$n$ junctions may be realized through subsequent annealing steps in different temperatures and atmospheres. This opens a new and more sustainable approach to engineering functionality in materials which can be upcycled through annealing. This would be far superior to making new materials with respect to energy consumption, time and environmental impact.

## ACKNOWLEDGEMENTS

D.R.S. acknowledges The Research council of Norway (FRINATEK Project No. 231430/20) and NTNU for financial support. S.P.S. and N.K. acknowledges The Research Council of Norway under the Nano2021 project (Project No. 228854) for financial support. The computational resources were provided by UNINETT Sigma2 (Projects No. NN9264k and ntnu243). D.M. acknowledges funding from the European Research Council (ERC) under the European Union's Horizon 2020 Research and Innovation Program (Grant Agreement No. 863691).

# Supplementary information:

# Controlling electronic properties of hexagonal manganites through aliovalent doping and thermoatmospheric history

Didrik R. Småbråten[1,2], Frida H. Danmo[1], Nikolai H. Gaukås[1], Sathya P. Singh[1], Nikola Kanas[1,3], Dennis Meier[1], Kjell Wiik[1], Mari-Ann Einarsrud[1], and Sverre M. Selbach[1,*]

[1]Department of Materials Science and Engineering, NTNU Norwegian University of Science and Technology, NO-7491 Trondheim, Norway

[2]Department of Sustainable Energy Technology, SINTEF Industry, PO Box 124 Blindern, NO-0314 Oslo, Norway.

[3]University of Novi Sad, BioSense Institute, 21000 Novi Sad, Serbia

*E-mail: selbach@ntnu.no

# Supplementary figures

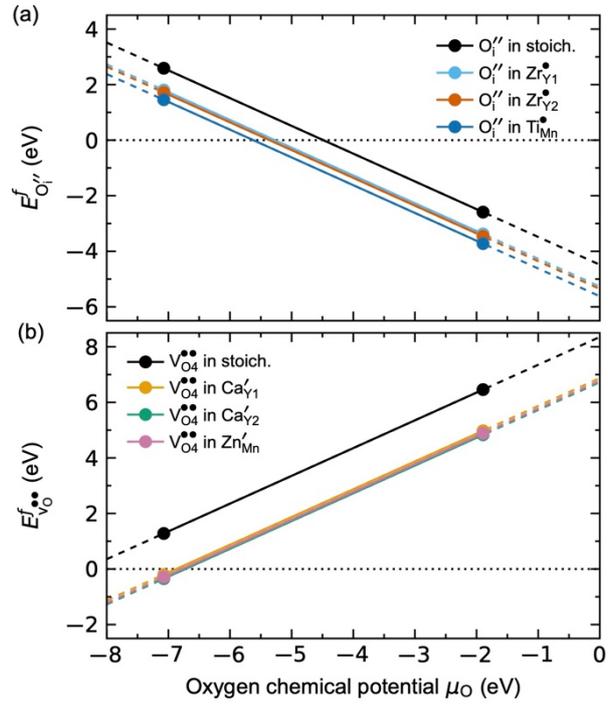

FIG. S1: A comparison of the calculated defect formation energies in 120 atoms 2x2x1 supercells as a function of oxygen chemical potential for (a) $O_i''$ in undoped and donor-doped $YMnO_{3+\delta}$, and (b) $V_O^{\bullet\bullet}$ in undoped and acceptor-doped $YMnO_{3-\delta}$.

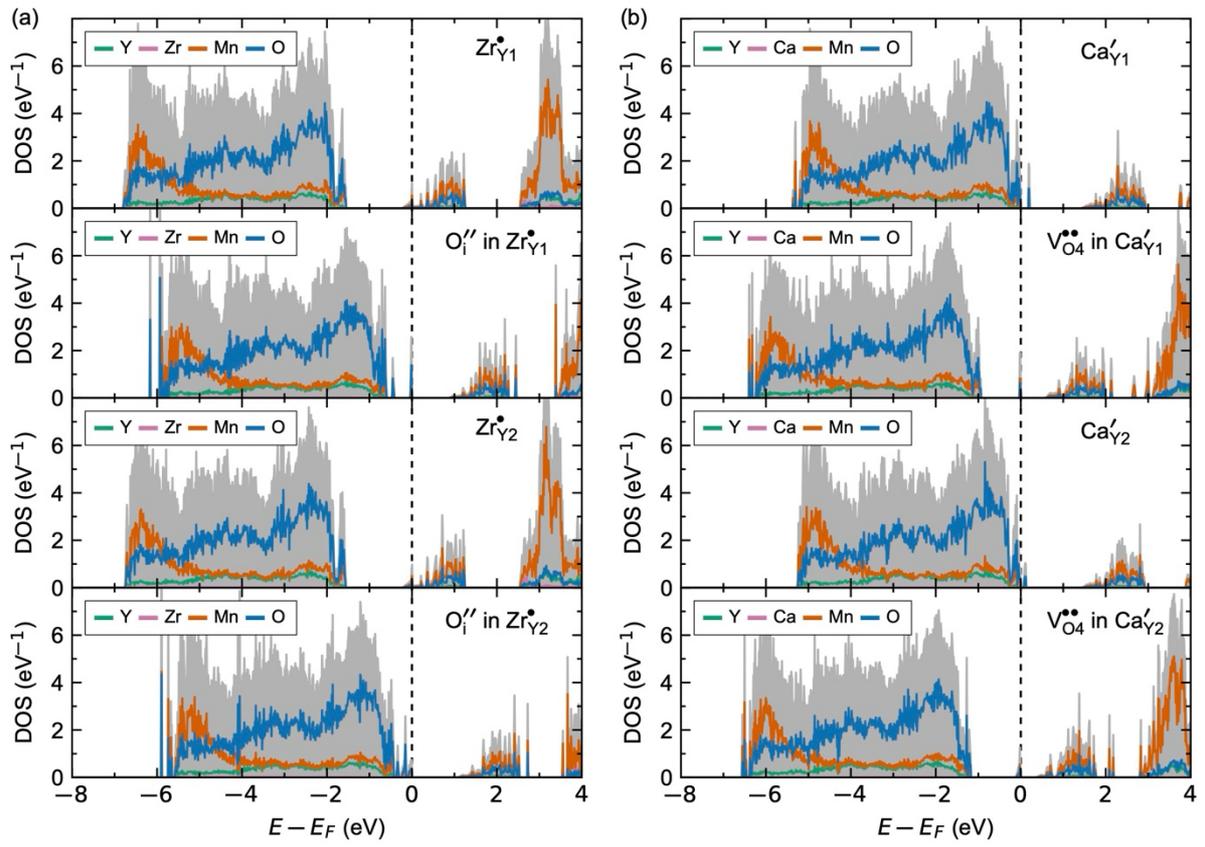

FIG. S2: Calculated DOS for 120 atoms 2x2x1 supercells with respect to Y substitution site for (a) Zr-doping with and without $O_i''$, and (b) Ca-doping with and without $V_O^{\bullet\bullet}$.

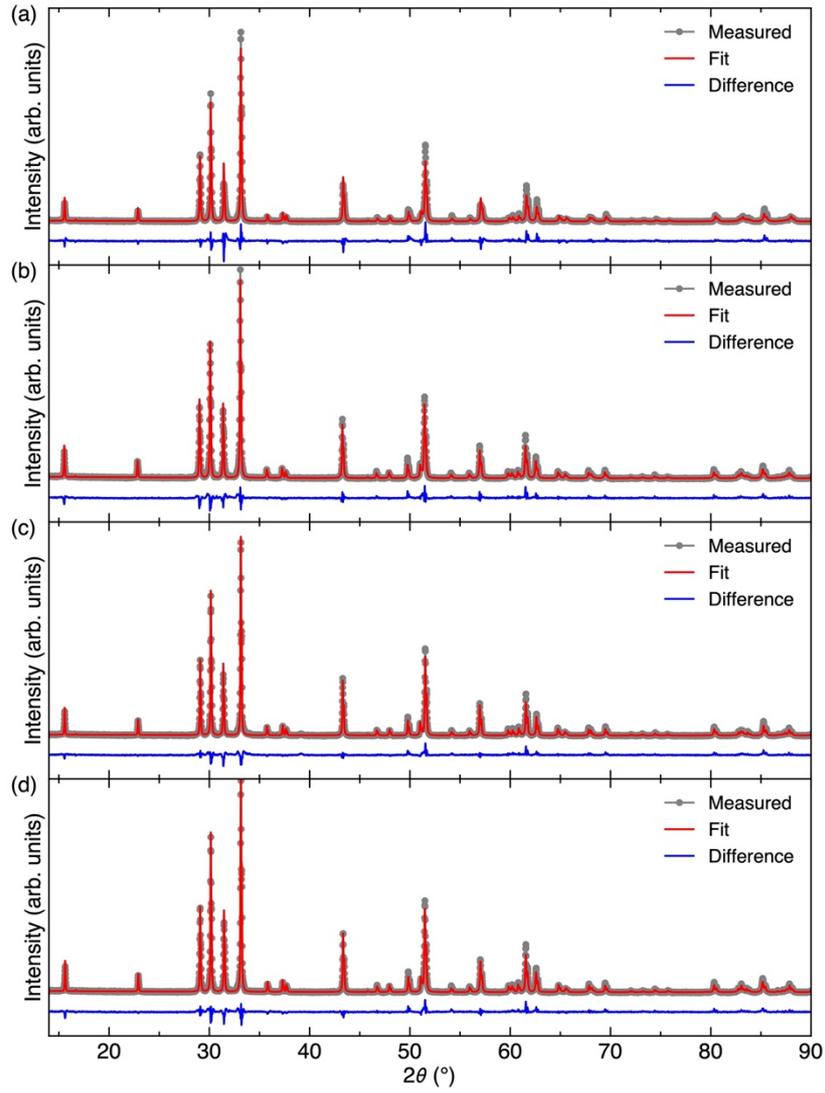

FIG. S3: XRD patterns and Rietveld refinements (space group $P6_3cm$) for as-prepared bulk powders of (a) Zr-, (b) Ti-, (c) Ca- and (d) Zn-doped YMnO$_3$. The measured patterns are shown in grey circles, the total fit in red, and the difference between the measured pattern and the fit in blue.

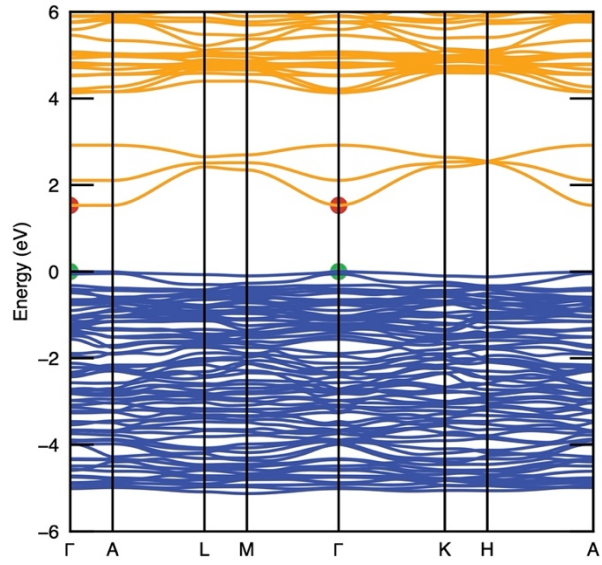

FIG. S4. Calculated electronic band structure of YMnO$_3$. The valence band maximum and conduction band minimum are illustrated by the green and red circles, respectively.

**Supplementary note 1: Comparison between Y1 and Y2 substitution**

A comparison of the density of states for Zr-doping on the two Y-sites with and without $O_i''$ is shown in FIG. S1(a), and for Ca-doping on the two Y-sites with and without $V_O^{\bullet\bullet}$ in FIG. S2(b). We find some qualitative differences in the density of states depending on which Y-site that is substituted, the general trends still remain. Hence, we expect similar conducting behavior, regardless of substitution site. Similarly, the calculated defect formation energies of $O_i''$ in donor-doped YMnO$_3$ [FIG. S1(a)], and $V_O^{\bullet\bullet}$ in acceptor-doped YMnO$_3$ [FIG. S1(b)], show similar energetics with respect to which Y-site that is substituted. We find a relative difference in $O_i''$ formation energies of 0.08 eV between Zr-doping on the two Y-sites, and a relative difference in $V_O^{\bullet\bullet}$ formation energies of 0.15 eV between Ca-doping on the two Y-site. These relative differences are insignificant as compared to the difference in formation energies between doped and undoped YMnO$_3$.

## Supplementary note 2: Defect chemistry model for Ti-doping

The annealing of Ti-doped YMnO$_3$ at intermediate temperatures in varying oxygen partial pressure can be described by the general chemical equilibrium

$$\tfrac{1}{2}Y_2O_3 + \tfrac{1-x}{2}Mn_2O_3 + xTiO_2 + \tfrac{\delta}{2}O_2(g) \leftrightarrow YMn_{1-x}Ti_xO_{3+\delta} + \tfrac{x}{4}O_2(g), \quad (S1)$$

where $x$ is the dopant concentration and $\delta$ is the oxygen hyper-stoichiometry. In sufficiently low $p(O_2)$, we can assume an oxygen stoichiometry equal to $\delta = 0$. Microscopically, Eq. (S1) becomes (in Kröger-Vink notation)

$$\tfrac{1}{2}Y_2O_3 + TiO_2 + Mn_{Mn}^{\times} \rightarrow Y_Y^{\times} + Ti_{Mn}^{\bullet} + Mn_{Mn}' + 3O_O^{\times} + \tfrac{1}{4}O_2(g). \quad (S2)$$

Here, the donor dopant $Ti_{Mn}^{\bullet}$ is charge compensated by reducing one Mn$^{3+}$ to Mn$^{2+}$ which renders the material *n*-type conducting, as expected for donor-doping. By increasing the oxygen partial pressure, oxygen interstitials start to form. Assuming an oxygen interstitial concentration according to $[O_i''] = \tfrac{1}{2}[Ti_{Mn}^{\bullet}]$, the oxygen interstitials are ionically charge compensated by the donor dopant. This is termed the neutral condition, described by

$$\tfrac{1}{2}Y_2O_3 + TiO_2 \rightarrow Y_Y + Ti_{Mn}^{\bullet} + 3O_O^{\times} + \tfrac{1}{2}O_i''. \quad (S3)$$

For these specific conditions, the material is expected to be neither *n*-type nor *p*-type dominating, which should give an electronic conductivity similar to that of stoichiometric YMnO$_3$ as the electronic charge carrier concentration remains unchained. Further increasing the oxygen content exceeding the neutral conditions, the excess oxygen interstitials are charge compensated by oxidizing Mn$^{3+}$ to Mn$^{4+}$ according to Eq. (5) in the main text, which renders the material *p*-type. In our DFT calculations, we add one oxygen interstitial per dopant, corresponding to the conditions $[O_i''] = [Ti_{Mn}^{\bullet}]$. Microscopically, Eq. (S1) becomes

$$\tfrac{1}{2}Y_2O_3 + TiO_2 + Mn_{Mn}^{\times} + \tfrac{1}{4}O_2(g) \rightarrow Y_Y^{\times} + Ti_{Mn}^{\bullet} + Mn_{Mn}^{\bullet} + 3O_O^{\times} + O_i''. \quad (S4)$$

The net result is one Mn$^{3+}$ being oxidized to Mn$^{4+}$, which renders the material *p*-type conducting.

## Supplementary note 3: Defect chemistry model for Zn-doping

The annealing of Zn-doped YMnO$_3$ at high temperatures in varying oxygen partial pressure can be described by the general chemical equilibrium

$$\tfrac{1}{2}Y_2O_3 + \tfrac{1-x}{2}Mn_2O_3 + ZnO + \tfrac{x}{4}O_2\,(g) \leftrightarrow YMn_{1-x}O_{3-\delta} + \tfrac{\delta}{2}O_2\,(g), \qquad (S5)$$

where $x$ is the dopant concentration and $\delta$ is the oxygen hypo-stoichiometry. In sufficiently high $p(O_2)$, we can assume an oxygen stoichiometry equal to $\delta = 0$. Microscopically, Eq. (S5) becomes

$$\tfrac{1}{2}Y_2O_3 + ZnO + Mn_{Mn}^{\times} + \tfrac{1}{4}O_2\,(g) \rightarrow Y_Y^{\times} + Zn_{Mn}' + Mn_{Mn}^{\bullet} + 3O_O^{\times}. \qquad (S6)$$

Here, the acceptor dopant $Zn_{Mn}'$ is charge compensated by oxidizing one Mn$^{3+}$ to Mn$^{4+}$ which renders the material $p$-type conducting, as expected for acceptor-doping. By decreasing the oxygen partial pressure, oxygen vacancies start to form. Assuming neutral conditions, corresponding to $[V_O^{\bullet\bullet}] = \tfrac{1}{2}[Zn_{Mn}']$, the oxygen vacancies are ionically charge compensated by the acceptor dopant according to

$$\tfrac{1}{2}Y_2O_3 + ZnO \rightarrow Y_Y^{\times} + Zn_{Mn}' + \tfrac{5}{2}O_O^{\times} + \tfrac{1}{2}V_O^{\bullet\bullet}. \qquad (S7)$$

For these specific conditions, the material is expected to be neither $n$-type nor $p$-type dominating, which should give an electronic conductivity similar to that of stoichiometric YMnO$_3$ as described for Ti-doping above. Further decreasing the oxygen content below neutral conditions, the excess oxygen vacancies are charge compensated by reducing Mn$^{3+}$ to Mn$^{2+}$ according to Eq. (6) in the main text, which renders the material $n$-type conducting. In our DFT calculations, we add one oxygen vacancy per acceptor dopant, corresponding to the conditions $[V_O^{\bullet\bullet}] = [Zn_{Mn}']$. Microscopically, Eq. (S5) becomes

$$\tfrac{1}{2}Y_2O_3 + ZnO + Mn_{Mn}^{\times} \rightarrow Y_Y^{\times} + Zn_{Mn}' + Mn_{Mn}' + 2O_O^{\times} + V_O^{\bullet\bullet} + \tfrac{1}{4}O_2\,(g), \qquad (S8)$$

where the net result is one Mn$^{3+}$ being reduced to Mn$^{2+}$, which renders the material $n$-type conducting.

# Supplementary note 4: Effective masses of holes and electrons in YMnO$_3$

Using parabolic fitting of the band edges of the electronic band structure in FIG. S4, we arrive at the estimated effective masses of holes and electrons summarized in Table S1.

TABLE S1. Estimated effective masses of holes and electrons at the valence band maximum and conduction band minimum, respectively, indicated in FIG. S4.

| Segment | Hole effective mass ($m_e$) | Electron effective mass ($m_e$) |
|---|---|---|
| Γ → A | 13.10 | Inf. |
| Γ → M | 1.13 | 0.59 |
| Γ → K | 3.40 | 0.60 |